\documentstyle[preprint,eqsecnum,aps,twoside]{revtex}

\newcommand{\be}{\begin{equation}}
\newcommand{\ee}{\end{equation}}
\newcommand{\bea}{\begin{eqnarray}}
\newcommand{\eea}{\end{eqnarray}}
\newcommand{\Slash}{\! \! \! /}

\newcommand{\SLAsh}{\! \! \! \! \! / \, \,}
\newcommand{\nn}{\nonumber}
\newcommand{\Punkt}{\quad .}
\newcommand{\Komma}{\quad ,}

\begin{document}

\preprint{MPG--VT--UR 81/96}

\title{ Critical Scattering and Two Photon Spectra \\
       for a Quark/Meson Plasma}

\author{P.~Rehberg}
\address{Institut f\"ur Theoretische Physik, Universit\"at Heidelberg, \\
         Philosophenweg 19, D--69120 Heidelberg, Germany}
\author{Yu.\,L.~Kalinovsky\thanks{Present address: {\it Joint Institute
        for Nuclear Research, Bogoliubov Theoretical Laboratory,
        141980, Dubna, Moscow region, Russia}}
        and D.~Blaschke}
\address{MPG Arbeitsgruppe ``Theoretische Vielteilchenphysik'', \\
         Universit\"at Rostock, D--18051 Rostock, Germany}

\maketitle
\begin{abstract}
At the Mott transition of a quark/meson plasma, mesons become unbound.
This leads to the effect of critical scattering, which is studied by
investigating photon pair production due to the process $q\bar
q\to\gamma\gamma$. This process can proceed either via direct
annihilation or via the formation of a mesonic resonance.  It is shown
that the latter channel leads to an enhancement of photon pairs with
invariant mass equal to the thermal pion mass.  The size of this effect
measures the time the temperature stays near the Mott temperature
during the evolution.  It is particularly pronounced for a first order
phase transition.  The $\pi^0\to\gamma\gamma$ decay gives a
strong background contribution and may make the observation of
critical scattering in two-photon spectra of present day heavy
ion collision experiments difficult. \\[2mm]
\end{abstract}

\section{Introduction} \label{introsec}
In ultrarelativistic heavy ion collision experiments at CERN SPS and
BNL AGS hadronic matter is produced under extreme conditions of
temperature and density. These conditions might be sufficient for the
transition to a new phase of matter, the quark gluon plasma
\cite{qm95,qm96}. Up to now, no unambiguous signature of this
hypothetical phase has been found, and it appears that a rather complex
analysis of observables has to be performed.  The inputs for such a
numerical analysis should be provided from effective approaches to QCD
in the nonperturbative low energy sector where chiral symmetry breaking
and confinement of QCD degrees of freedom are the prominent features.

Although plagued with nonrenormalizability and the absence of
confinement, the Nambu-Jona-Lasinio (NJL) model
\cite{volkov,weise,sandi,hatkun} has been developed recently beyond the
applications for calculation of hadronic mass spectra towards a kinetic
description of hadronic matter at the quark-hadron phase transition
\cite{kinetik,su3hadron}.  A particularly interesting problem within
this context is the description of hadrons as bound states of quarks
and antiquarks and their dissolution at finite temperature and density
due to thermal excitation or to compression, i.\,e. the Mott effect
\cite{gerry}.  At temperatures below the Mott temperature $T_M$, the
model contains massive constituent quarks and bound state mesons as
degrees of freedom.  For temperatures above $T_M$, these bound states
dissolve and become resonant (nonperturbative) correlations. Whereas
the appearance of free quarks at $T<T_M$ is an artifact of the NJL
model, the situation for $T\ge T_M$ is fairly realistic, except for the
absence of gluons.  We thus apply the model for the description of
mesonic correlations at $T\ge T_M$.  It has been argued elsewhere
\cite{gerry,pihad} that the Mott transition leads to the effect of
critical scattering and may have experimental consequences.  In this
work, we investigate the influence of such correlations in a quark
plasma on the two-photon spectrum which is one of the observables
measured in nucleus-nucleus collisions, see e.\,g. \cite{wa80}, and
which has been previously studied in Refs.~\cite{Redlich,ThePeople}.
To this end, we present a model study of the process $q \bar q \to
\gamma \gamma$ employing the two flavor NJL model. In the $s$-channel,
this process proceeds via the formation of a virtual meson. At the Mott
temperature, this meson becomes unbound, which leads to a divergent
scattering length. We estimate the magnitude of this critical
scattering effect and its sensitivity to the order of the phase
transition by comparing the spectra containing resonance contributions
with those, which only contain contributions of direct annihilation.
However, we give only a crude estimate of the contributions from
$\pi^0$ decay into photons after freezeout.

Since presently no consistent transport theory modeling the chiral
phase transition is available, we apply a hydrodynamical scenario in
order to obtain the total photon spectra. In this approach, we use a
model equation of state \cite{kohl,rischke} which interpolates smoothly
between a first and a second order phase transition.  The results of
our investigation show that the yield of photon pairs with invariant
mass $m_\pi(T_M)$ depends strongly on the time the temperature stays
near the critical temperature.

This paper is structured as follows: The calculation of photon production
cross sections including resonance contributions is performed
in Section~\ref{crossec}. In Section~\ref{hydro} we apply the Bjorken
hydrodynamical expansion scenario \cite{Bj} in order to calculate the
invariant mass spectrum of photon pairs from a hadronizing quark
plasma. In Section~\ref{end}, we summarize and present the conclusions.

\section{Photon Production Cross Sections} 
\label{crossec}
\subsection{Review of Formalism}
For the calculation of photon production cross sections we employ the
Nambu--Jona-Lasinio (NJL) model in its $SU_f(2)$ version\cite{sandi,hatkun}.
The Lagrangian for this model reads
\be
{\cal L} = \bar\psi(i\partial\Slash - m_{0q}) \psi
         + G\left[ (\bar\psi\psi)^2 + (\bar\psi i\gamma_5\vec\tau\psi)^2\right]
\Punkt \label{lagra}
\ee
Here $G$ is a coupling constant of dimension MeV$^{-2}$ and $\vec\tau$ the
Pauli matrices in flavor space. The quark wave function $\psi$ carries
flavor and color degrees of freedom, which are implicitly summed in
Eq.~(\ref{lagra}). The symbol $m_{0q}$ denotes the current quark mass,
which explicitly breaks chiral invariance.

The masses of the constituent quarks are computed in leading order in an
expansion in the inverse number of colors, $1/N_c$ \cite{expand}. For the
constituent quark masses, this corresponds to a Hartree
approximation. One obtains the gap equation for the physical masses
\be
m_q = m_{0q}+4iGN_c\mbox{tr}_\gamma S(x,x)
    = m_{0q} - \frac{GN_c}{\pi^2} m_q A(m_q) \Komma \label{gapeq}
\ee
where the function $A(m_q)$ is given in the imaginary time Matsubara
formalism \cite{fetwal} as
\be
A(m_q) = \frac{16\pi^2}{\beta} \sum_n e^{i\omega_n\eta}
\int\limits_{|\vec p| < \Lambda} \frac{d^3p}{(2\pi)^3}
\frac{1}{(i\omega_n)^2-E^2} \Punkt
\ee
Here we have abbreviated $E=\sqrt{\vec p^2 + m_q^2}$, and
$\omega_n=(2n+1)\pi/\beta$ is the fermionic Matsubara frequency. The
symbol $\beta=1/T$ denotes the inverse temperature. The function
$A(m_q)$ is given explicitly by
\cite{loopies}
\be
A(m_q) = -4 \int_0^\Lambda dp \frac{p^2}{E} \tanh(\beta E/2) \Komma
\ee
where we have used an $O(3)$ cutoff to make the integral finite.

Mesonic correlations in the pseudoscalar (pion) and scalar (sigma) channels 
of the $q \bar q$ interaction are treated in the random phase approximation
\cite{sandi}. This leads to the form
\be
{\cal D_\pi}(k_0, \vec k) = \frac{2G}{1-4G\Pi^P(k_0, \vec k)}
\label{piprop}
\ee
for the pion propagator and
\be
{\cal D_\sigma}(k_0, \vec k) = \frac{2G}{1-4G\Pi^S(k_0, \vec k)} \Komma
\label{sigprop}
\ee
for the $\sigma$ propagator. In Eqs.~(\ref{piprop}) and (\ref{sigprop}),
$\Pi^P$ and $\Pi^S$ denote the irreducible pseudoscalar and scalar 
polarization functions, respectively. They are given explicitly by
\cite{su3hadron}
\bea
\Pi^P(k_0, \vec k) &=& -\frac{N_c}{8\pi^2}\left[2A(m_q)-(k_0^2-\vec k^2)
                       B_0(k_0, \vec k)\right] \\
\Pi^S(k_0, \vec k) &=& -\frac{N_c}{8\pi^2}\left[2A(m_q)+(4m_q^2-k_0^2+\vec k^2)
                       B_0(k_0, \vec k)\right] \Komma
\eea
where the function $B_0(k_0, \vec k)$ is defined as the analytical
continuation of
\be
B_0(i\nu_m, \vec k) = \frac{16\pi^2}{\beta} \sum_n e^{i\omega_n\eta}
\int\limits_{|\vec p| < \Lambda} \frac{d^3p}{(2\pi)^3}
\frac{1}{[(i\omega_n)^2-E^2]} \frac{1}{[(i\omega_n-i\nu_m)^2-E'^2]}
\label{B0def}
\ee
to real values of $i\nu_m\to k_0$. In Eq.~(\ref{B0def}), we have used the 
notation
$E'=\sqrt{(\vec p - \vec k)^2 + m_q^2}$. In the following, we will need
the function $B_0$ for the special case $\vec k = 0$, where it can be
expressed as \cite{loopies}
\bea
B_0(k_0, \vec 0) &=& 8 {\cal P} \! \! \! \int_0^\Lambda dp
\frac{p^2\tanh(\beta E / 2)}{E[4E^2-k_0^2]}
\\ \nonumber &+&
i\pi\sqrt{1-\frac{4m_q^2}{k_0^2}} \tanh\left(\frac{\beta k_0}{4}\right)
\Theta(k_0-2m_q)\Theta\left(2\sqrt{\Lambda^2+m_q^2}-k_0\right) \Punkt
\eea
One can immediately see from this form, that $B_0(k_0,\vec 0)$ is a
real function for $k_0<2m_q$ and a complex function for $k_0>2m_q$.

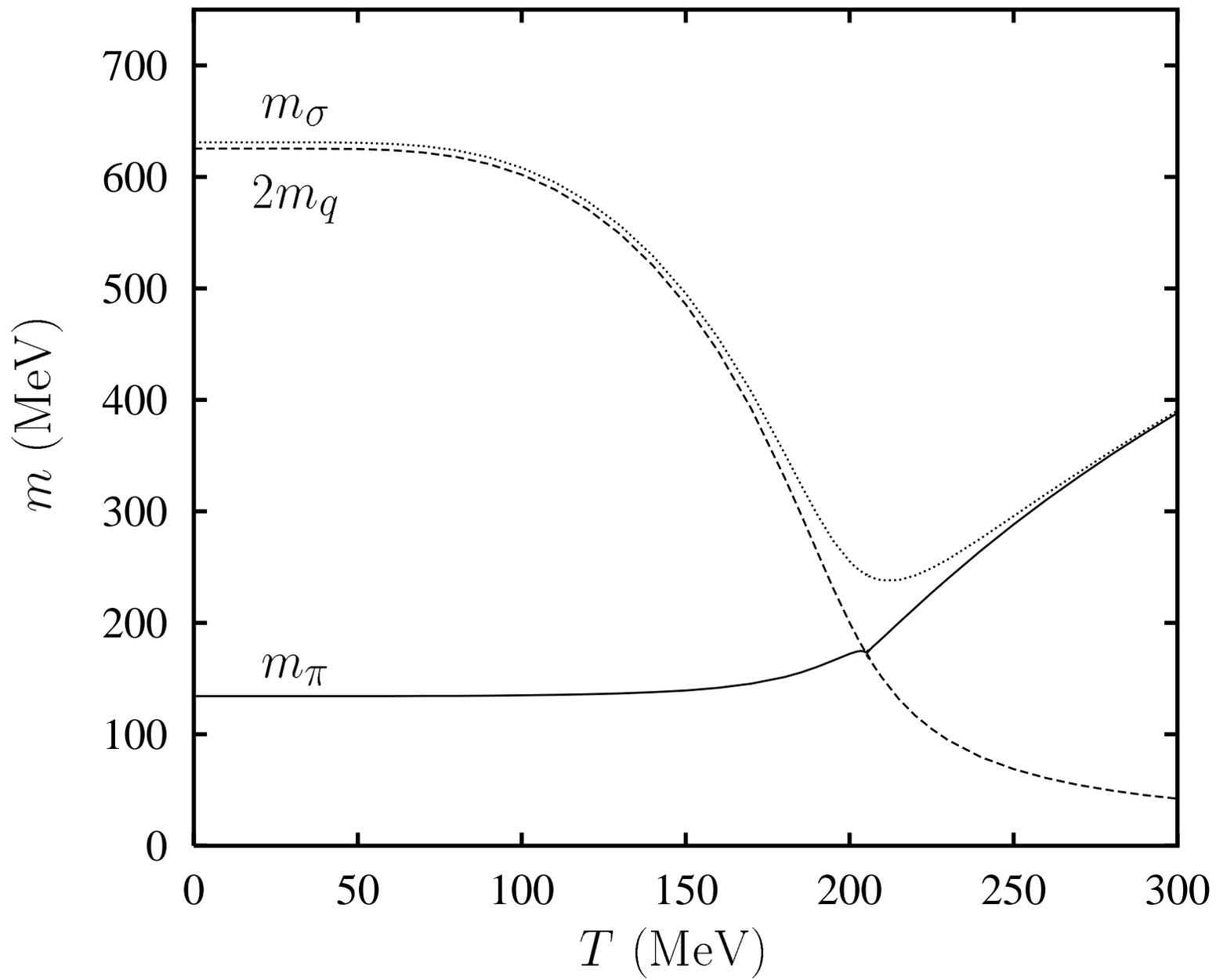
\begin{figure}
\caption[]{Mass spectrum of the $SU_f(2)$ NJL model. Solid line: pion mass,
          dashed line: double quark mass, dotted line: sigma mass.}
\label{masses}
\end{figure}

The masses of the $\pi$ and $\sigma$ mesons are obtained via the
dispersion relations \cite{sandi}
\bea
1-4G\Pi^P(\tilde m_\pi, \vec 0) &=& 0 \label{pidis} \\
1-4G\Pi^s(\tilde m_\sigma, \vec 0) &=& 0 \label{sigdis} \Punkt
\eea
Note that these are in general complex equations and the solution may
contain a nonzero width:
\be
\tilde m_\alpha = m_\alpha - \frac{i}{2}\Gamma_\alpha \quad, 
\qquad \alpha = \pi, \sigma \Punkt
\ee
In this case, the function $B_0(k_0,\vec 0)$ has to be analytically
continued to complex values of $k_0$ \cite{gerry}.  However,
Eq.~(\ref{pidis}) at low temperatures has {\em real} (bound state)
solutions for $\tilde m_\pi$ with $m_\pi(T) < 2m_q(T)$. At the Mott
temperature $T_M$, one obtains $m_\pi(T_M)=2m_q(T_M)$. For higher
temperatures, the pion becomes a resonant state. Further details about
this transition can be found in Ref.~\cite{gerry}. The sigma meson
differs from the pion in that it is a resonance for all temperatures.
In Fig.~\ref{masses}, we show the result of a numerical solution of
Eqs.~(\ref{gapeq}), (\ref{pidis}) and (\ref{sigdis}), using the
parameter set $m_{0q}=5.0~\mbox{MeV}$, $\Lambda=653.3~\mbox{MeV}$ and
$G\Lambda^2=2.105$, as a function of temperature.  At low temperatures,
the pion mass, which is indicated by the solid line, lies below the
mass of its constituents, which is given by the dashed line. At
$T=T_M$, one obtains $m_\pi(T_M)=2m_q(T_M)$, and the pion moves from
being a bound state to a resonant state. For our parameter set, this
happens at $T_M=205~\mbox{MeV}$. The mass of the sigma meson, which is
indicated by the dotted line in Fig.~\ref{masses}, is larger than
$2m_q$ in the whole temperature range. The sigma meson is thus a
resonant state at all temperatures.

\subsection{Photon Production Cross Sections}
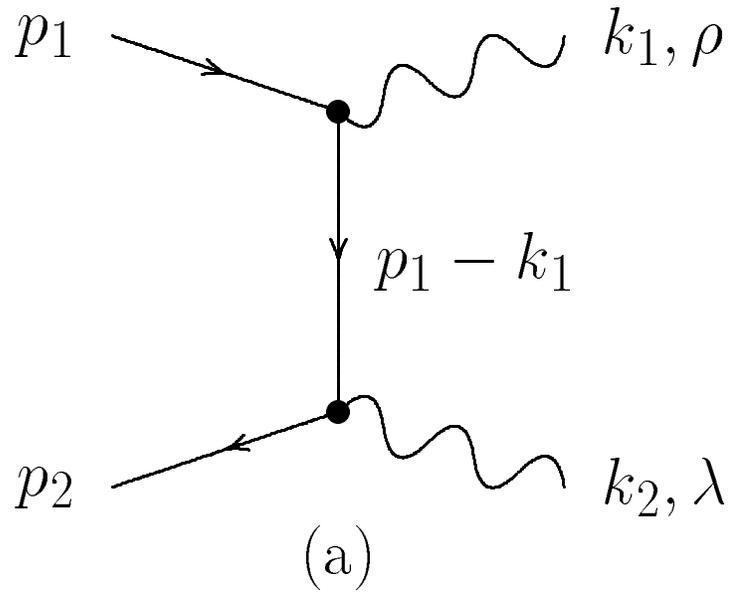
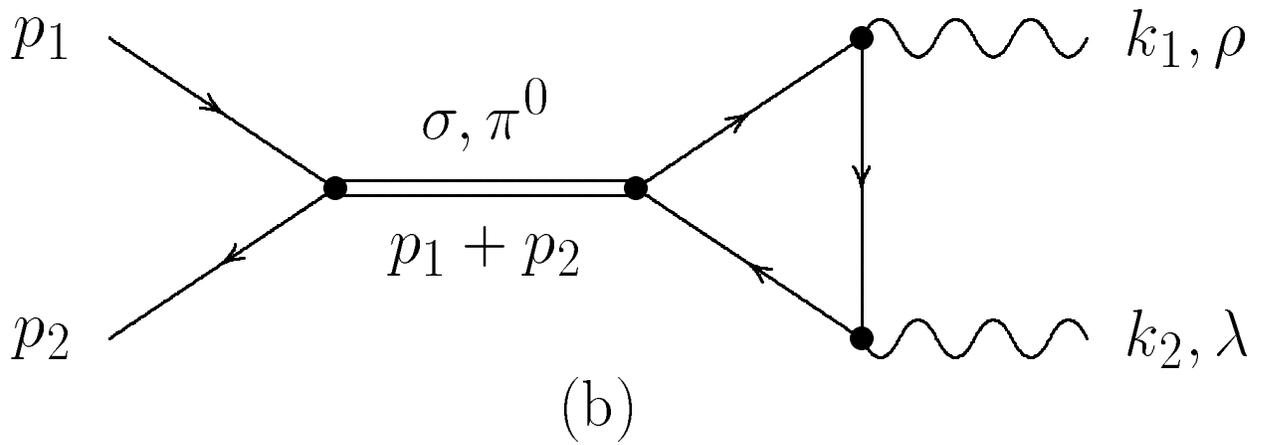
\begin{figure}
\caption[]{Feynman diagrams for $q\bar q\to\gamma\gamma$. Solid lines
	   denote quarks, wavy lines photons and double lines mesons.
           Graph (a) shows the direct diagram, graph (b) the $s$--channel
           diagrams. Two more diagrams, which arise by the substitution
           $k_1\leftrightarrow k_2$, $\rho \leftrightarrow \lambda$
           are not shown.}
\label{feyndiag}
\end{figure}

The Feynman diagrams, which contribute to the process $q\bar
q\to\gamma\gamma$ in lowest order in $1/N_c$ are shown in
Fig.~\ref{feyndiag}.  Beside the direct terms, i.\,e. those which do
not contain resonance contributions and which are already
present in quantum electrodynamics \cite{lanlif}, there are four
diagrams, in which the incoming quarks first form a virtual $\pi^0$ or
$\sigma$ state, which afterwards decays into two photons. We write the
transition amplitude for the $t$--channel direct term shown in
Fig.~\ref{feyndiag}a as
\be
-i{\cal M}_t = \delta_{c_1,c_2} \bar v_2(p_2) (-iq_fe\epsilon_2\SLAsh)
               \frac{i}{p_1 \SLAsh - k_1 \SLAsh - m_q}
               (-iq_fe\epsilon_1\SLAsh) u(p_1)
\Punkt \label{tchan}
\ee
In this equation, $p_1$ and $p_2$ denote the four-momenta of the
incoming quarks, $k_1$ and $k_2$ the four-momenta of the outgoing
photons and $\epsilon_1$ and $\epsilon_2$ their respective polarization
vectors. The electron charge magnitude is denoted by $e$, $q_f$ is the
quark charge factor; one has $q_u=+2/3$ for up quarks and
$q_d=-1/3$ for down quarks. The Kronecker symbol
$\delta_{c_1,c_2}$ guarantees that only color singlet states can
annihilate into two photons. The transition amplitude for the
$u$--channel direct graph is evaluated analogously to be
\be
-i{\cal M}_u = \delta_{c_1,c_2} \bar v_2(p_2) (-iq_fe\epsilon_1\SLAsh)
               \frac{i}{p_1 \SLAsh - k_2 \SLAsh - m_q}
               (-iq_fe\epsilon_2\SLAsh) u(p_1)
\Punkt \label{uchan}
\ee
Multiple rescattering in the initial state leads to the formation of a 
strong correlation in the s-channel as is shown diagrammatically in 
Fig.~\ref{feyndiag}b where the correlation is described by a meson propagator.
The transition amplitude corresponding to sigma propagation in the
$s$--channel can be written as
\be
-i{\cal M}_{s,\sigma} = \delta_{c_1,c_2} \bar v(p_2) u(p_1)
              \left[i{\cal D}_\sigma(\sqrt{s},\vec 0)\right]
              \left[-ia(s)(k_1^\lambda k_2^\rho-g^{\rho\lambda}k_1k_2)\right]
              \epsilon_{1\rho}\epsilon_{2\lambda}
\Punkt \label{s1chan}
\ee
In this expression, we have restricted ourselves to the kinematical
situation that the center of mass system rests with respect to the
medium. The factor $-ia(s)(k_1^\lambda k_2^\rho-g^{\rho\lambda}k_1k_2)$
stems from the triangle part of the diagram in Fig.~\ref{feyndiag}b.
The explicit form of $a(s)$ is derived in Appendix~\ref{appvertex}.
Here we only note that $a(s)$ contains a factor $(1-4m_q^2/s)$, which
leads to a suppression of this diagram near threshold. The transition
amplitude corresponding to the propagation of a pionic mode is given by
\be
-i{\cal M}_{s,\pi} = \delta_{c_1,c_2} \bar v(p_2)
                     (i\gamma_5\tau_{ff}^3) u(p_1)
                     \left[i{\cal D}_\pi(\sqrt{s},\vec 0)\right]
                     \left[-ib(s)
                        \varepsilon^{\mu\nu}_{\phantom{\mu\nu}\lambda\rho}
                        k_1^\lambda k_2^\rho \right]
                     \epsilon_{1\mu}\epsilon_{2\nu}
\Komma \label{s2chan}
\ee
where the triangle contributions are contained in the factor
$-ib(s)\varepsilon^{\mu\nu}_{\phantom{\mu\nu}\lambda\rho}k_1^\lambda
k_2^\rho$. Here, $\varepsilon$ denotes the totally antisymmetric
tensor. As was the case for $a(s)$, an explicit expression for $b(s)$
is given in Appendix~\ref{appvertex}. Compared to Eq.~(\ref{s1chan}),
this expression also contains an additional isospin factor
$\tau_{ff}^3$, which is equal to $+1$ for up quarks, and $-1$ for down
quarks in the initial state.

The integrated cross section is given by
\be
\sigma_{q\bar q\to\gamma\gamma}(s,T) = \frac{1}{16\pi s(s-4m_q^2)}
                  \int_{t_{\rm min}}^{t_{\rm max}} dt
                  \overline{|{\cal M}|^2} \Komma
\label{sigtot}
\ee
where
\be
\overline{|{\cal M}|^2} = \frac{1}{4N_c^2} \sum_{s,c}
|{\cal M}_t + {\cal M}_u + {\cal M}_{s,\sigma} + {\cal M}_{s,\pi} |^2
\ee
is the spin and color averaged square of the transition amplitude and
the integration limits are given by
\be
t_{\rm min} = m_q^2 - \frac{s}{2} \hspace{1cm}
t_{\rm max} = t_{\rm min} + \frac{1}{2} \sqrt{s(s-4m_q^2)} \Punkt
\ee
Explicit expressions for $\overline{|{\cal M}|^2}$ and its integral
over $t$ can be found in Appendix~\ref{appmatel}.

At the Mott temperature, the pion appearing in the $s$--channel graph
of Fig.~\ref{feyndiag}(b) switches from a bound to a resonant state.
It is a well known fact from scattering theory, that a
bound state at threshold leads to a divergent scattering length. In the
present calculation, this can be shown explicitly by noticing that the
pion propagator ${\cal D}_\pi$ contains a singularity at $T=T_M$.
The leading behavior of ${\cal D}_\pi$ can be obtained from the
observation that $B_0(\sqrt{s},\vec 0)$ can be expanded in a power
series in $(4m_q^2-s)^{1/2}$ \cite{gerry}:
\be
B_0(\sqrt{s}, \vec 0) = {\cal A} + {\cal B} (4m_q^2-s)^{1/2}
+ {\cal C} (4m_q^2-s) + {\cal O}((4m_q^2-s)^{3/2}) \Punkt
\ee
The pion propagator can thus near threshold be approximated by
\be
{\cal D}_\pi(\sqrt{s}, \vec 0) =
\frac{2G}{\left[ 1+\xi\left( A(m_q)
-2m_q^2 {\cal A}\right)\right] - \frac{\xi}{2} \left[
s{\cal B}(4m_q^2-s)^{1/2} -({\cal A} + s{\cal C})(4m_q^2-s) \right]}
\label{propex}
\ee
with $\xi = (N_cG)/\pi^2$.
Since one has $m_\pi=2m_q$ at $T=T_M$, the first square bracket in
Eq.~(\ref{propex}) has to vanish at this temperature and one obtains
\be
{\cal D}_\pi(\sqrt{s}, \vec 0) \sim (s-4m_q^2)^{-1/2} \Punkt
\ee
Via Eqs.~(\ref{litschi}) and (\ref{sigtot}), one obtains
\be
\sigma_{q\bar q\to\gamma\gamma} \sim (s-4m_q^2)^{-3/2} \label{sigkri}
\ee
near threshold. The pion propagator diagram thus leads to a strong
enhancement of the production of soft photons at the Mott transition
when compared with the direct contributions. Due to the temperature
dependence of the pion mass (see Fig. \ref{masses}) the peak in the
two-photon invariant mass spectrum at $\sqrt{s}=m_\pi$ is shifted
upwards to a value of $m_\pi(T_M)=173$~MeV. Note that the effect of a
rising pion mass at high temperatures is also visible in other
effective models \cite{sepsig,confin} and in lattice calculations
\cite{laer}. Moreover, the $\pi^0 \to \gamma \gamma$ decay rate is
strongly enhanced at the Mott transition which is an effect akin to the
phenomenon of critical opalescence known from solid state physics
\cite{gerry}. Numerically, one finds that the resonance contributions
are most pronounced near the Mott temperature, while their importance is
strongly reduced compared to the direct terms at higher temperatures. A
similar behaviour has been found previously in Refs.~\cite{sapho,anomal}.

\section{Hydrodynamical Expansion}
\label{hydro}
The cross sections calculated in Sec.~\ref{crossec} are not measureable
directly in heavy ion experiments. Instead, one has to integrate the
production rates over the evolution of the system. A crude model for
the expansion and cooling of the plasma has been given by Bjorken
\cite{Bj}. Within this model, we will use different {\em ans\"atze\/}
for the equation of state and investigate their influence on the
physical spectra.

\subsection{Production Rates per Space--Time Element}
The production rate of photon pairs with invariant mass $M$ per
unit space--time volume is given by
\bea
\frac{dN_{\gamma\gamma}}{d^4xdM}&=&2M\int\frac{d^3p_1}{(2\pi)^3}
\frac{d^3p_2}{(2\pi)^3} v_{\rm rel} \sigma_{\gamma\gamma}(M^2,T)
\label{prate1} \\ \nn & & \hspace{2cm} \times
\left[2N_c f_F(\beta p_1^\mu u_\mu)\right]
\left[2N_c f_F(\beta p_2^\mu u_\mu)\right]
\delta(M^2-(p_1+p_2)^2)\Komma
\eea
where $f_F(x)=1/(\exp(x)+1)$ is the Fermi distribution function and
$\sigma_{\gamma\gamma}$ the sum of the cross sections for $u\bar
u\to\gamma\gamma$ and $d\bar d\to\gamma\gamma$. As in
Section~\ref{crossec}, we neglect effects that stem from a relative
motion of the quark pair with respect to the rest frame of the medium.
In principle, Eq.~(\ref{prate1}) should be supplemented by Bose
enhancement factors, which account for the stimulated emission of
photons in the medium.  Since the form of such enhancement factors is
unknown and can only be determined in a nonequilibrium theory, we
discard any such factors \cite{Redlich,ThePeople}, assuming that the
photon density stays low due to the short lifetime of the plasma and
thus prohibiting equilibration.  Note that the inclusion of an
enhancement factor, even at the finally equilibrated level, only alters
our results quantitatively by at most a factor of 10.  Our qualitative
picture remains however unchanged.  The fluid velocity, which has to be
supplied by the expansion model, is denoted by $u_\mu$.  As it turns
out, the final result does not depend on $u_\mu$, so one can evaluate
(\ref{prate1}) under the assumption that the fluid is at rest.
The final result for the production rates reads \cite{Redlich,Dom}
\be
\frac{dN_{\gamma\gamma}}{d^4xdM}=
\frac{(2N_c)^2}{(2\pi)^4}
\frac{M^2\sqrt{M^2-4m_q^2}\sigma_{\gamma\gamma}(M^2,T)}{\beta}
\int_{m_q}^\infty dE f_F(\beta E)\log\frac{1+\exp(-\beta E_-)}
{1+\exp(-\beta E_+)} \label{prate2}
\ee
with
\be
E_\pm = \frac{1}{2m_q^2}\left[ (M^2-2m_q^2)E \pm
\sqrt{M^2(M^2-4m_q^2)(E^2-m_q^2)} \right] \Punkt
\ee
An expansion of Eq.~(\ref{prate2}) near threshold shows that the
singular part of the production rate behaves as
\be
\frac{dN_{\gamma\gamma}}{d^4xdM}\sim (M^2-4m_q^2) \sigma_{\gamma\gamma}(M^2,T)
\Punkt
\ee
At $T\ne T_M$, where $\sigma_{\gamma\gamma}$ displays only kinematical
singularities, the production rate thus vanishes at threshold. At the
Mott temperature, however, one obtains from Eq.~(\ref{sigkri}), that
the production rate diverges as
\be
\frac{dN_{\gamma\gamma}}{d^4xdM}\sim (M^2-4m_q^2)^{-1/2} \label{nkri}
\ee
near threshold. Note that this is an integrable singularity, i.\,e.
although the production rate contains a divergence, the total number of
photons produced stays finite.

\subsection{Expansion Model}
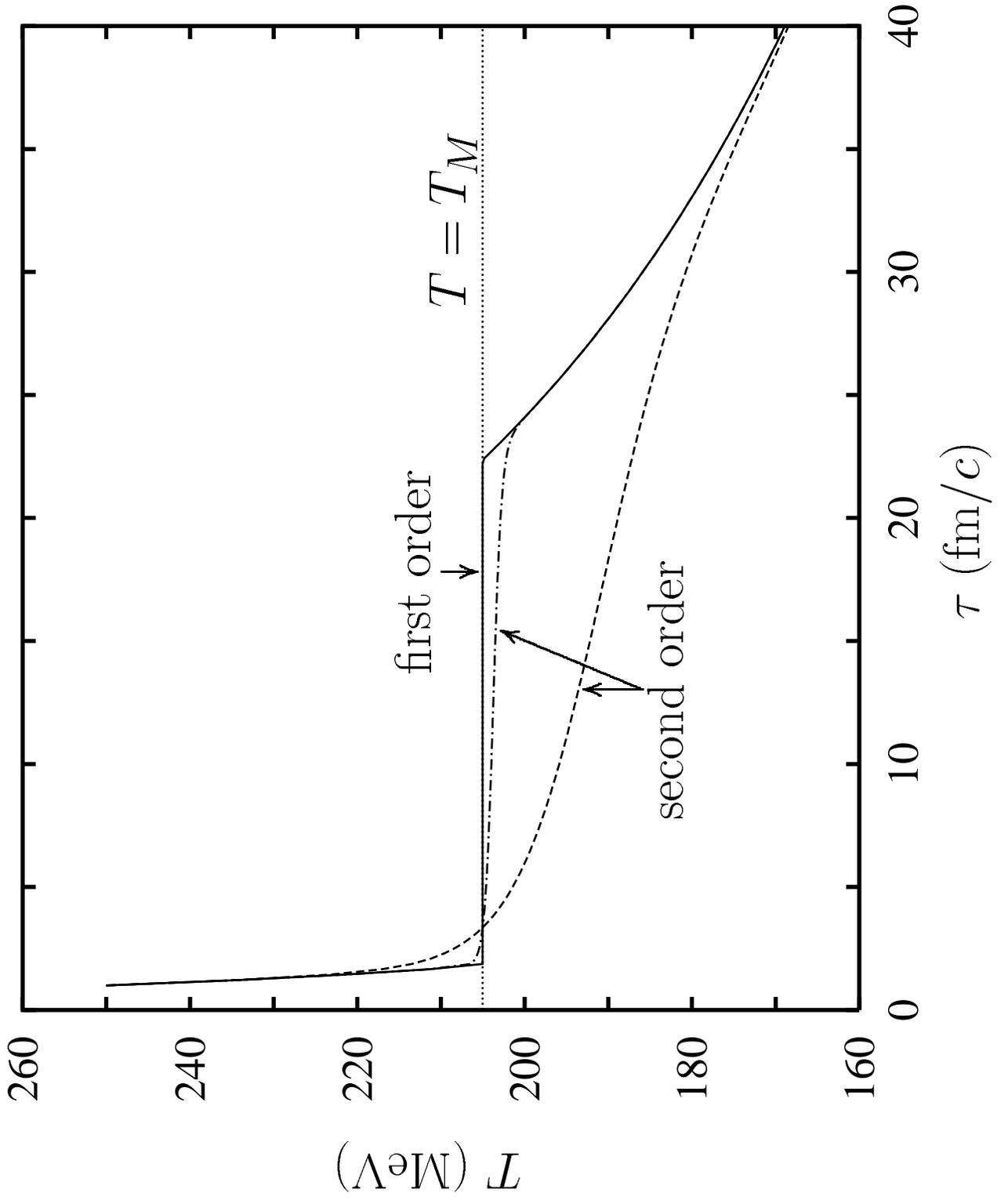
\begin{figure}
\caption[]{Temperature evolution for the parameter set $T_0=250~\mbox{MeV}$
           and $\tau_0=1~\mbox{fm}/c$. The solid line shows the first order
	   scenario, the dashed and dot--dashed curves the second order
	   scenario for $\Delta T=10~\mbox{MeV}$ and $\Delta
	   T=1~\mbox{MeV}$, respectively. The dotted line indicates the
	   Mott temperature.}
\label{temper}
\end{figure}

As a simple model for the hydrodynamical expansion, we employ Bjorkens
one dimensional model \cite{Bj}. The essence of this model is contained
in the two equations for the fluid velocity $u_\mu$
\be
u_\mu = \frac{1}{\tau} (t, 0, 0, z) \label {bjvel}
\ee
and the entropy density $s$
\be
\frac{s(\tau)}{s(\tau_0)} = \frac{\tau_0}{\tau} \Komma \label{bjent}
\ee
where $\tau=\sqrt{t^2-z^2}$ is the proper time of the fluid element.
Note that all thermodynamic quantities depend on $\tau$ only.  In
Eq.~(\ref{bjent}), $\tau_0$ denotes the formation time of the plasma,
which we treat as a free parameter in the following.  To compute the
temperature distribution from Eq.~(\ref{bjent}), one has to make
assumptions about the equation of state of the plasma. In the
following, we will compare two different scenarios:

(i) {\em First order phase transition:\/} The entropy density changes
discontinously at $T=T_M$. In this case, the equation of state reads
\cite{kohl,Redlich}
\be
s = \frac{4\pi^2}{90} T^3 \times \left\{
\begin{array}{ccc}
d_Q & \qquad \mbox{if} \qquad & T > T_M \\
f d_Q + (1-f) d_H & \qquad \mbox{if} \qquad & T = T_M \\
d_H & \qquad \mbox{if} \qquad & T < T_M
\end{array} \right. \Punkt \label{firsteos}
\ee
The parameter $f$  in Eq.~(\ref{firsteos}) is the volume fraction of
the system, which is in the quark-gluon phase, $(1-f)$ the volume
fraction, which is in the hadron phase.  The symbols $d_i$ denote the
effective numbers of degrees of freedom; one has $d_Q=37$, $d_H=3$.
From Eq.~(\ref{firsteos}), one obtains the following behavior of
temperature:
\begin{enumerate}
\item The expansion starts at $\tau=\tau_0$ with an initial temperature
      $T_0$. Afterwards, the temperature drops according to
      \be
      T = T_0 \left(\frac{\tau_0}{\tau}\right) ^ {1/3} \Komma
      \ee
      until it reaches the Mott temperature at time
      \be
      \tau_1 = \tau_0 \left(\frac{T_0}{T_M}\right)^3 \Komma
      \ee
      at which the mixed phase begins.
\item The second stage of the expansion is isothermal with $T=T_M$. During
      this stage, the relative volume of the plasma evolves according to
      \be
      f(\tau) = \frac{1}{d_Q-d_H}\left(d_Q\frac{\tau_1}{\tau}-d_H\right) \Punkt
      \ee
      This stage ends at time
      \be
      \tau_2=\frac{d_Q}{d_H} \tau_1 \Komma
      \ee
      at which $f(\tau_2)=0$ and the system enters the purely hadronic
      phase.
\item During the third stage, the temperature of the system drops as
      \be
      T = T_M \left(\frac{\tau_2}{\tau}\right) ^ {1/3} \Punkt
      \ee
      This stage lasts until breakup.
\end{enumerate}
The solid curve of Fig.~\ref{temper} gives a specific example for
the parameters $T_0=250~\mbox{MeV}$ and $\tau_0=1~\mbox{fm}/c$.
The temperature stays constant from $\tau_1=1.8~\mbox{fm}/c$
to $\tau_2=22.4~\mbox{fm}/c$.

The physical photon spectra have to be obtained by a space--time
integration of the production rates.  In the Bjorken model, the
integration over the transverse space directions simply contributes a
factor $\pi R_A^2$, where $R_A$ is the radius of the colliding nuclei.
Since the thermodynamic quantities only depend on $\tau$, it is useful
to switch to the integration variables $\tau$ and
\be
\eta = \frac{1}{2} \log \frac{t+z}{t-z} \Punkt
\ee
The final result is
\be
\frac{dN_{\gamma\gamma}}{dMd\eta} = \pi R_A^2 \left(
\int_{\tau_0}^{\tau_1} d\tau \, \tau \frac{dN_{\gamma\gamma}}{d^4xdM} +
\int_{\tau_1}^{\tau_2} d\tau \, \tau f(\tau) \frac{dN_{\gamma\gamma}}{d^4xdM}
\right) \Komma \label{firstspect}
\ee
which gives the number of photons produced in the central rapidity
interval of the collision. Note that Eq.~(\ref{firstspect}) contains
only contributions from the plasma phase. Although the NJL model would
allow for $q\bar q$ annihilation in the hadronic phase, we discard
these unphysical contributions by restricting the integration limits.

(ii) {\em Second order phase transition:\/} For a second order phase
transition, the jump in Eq.~(\ref{firsteos}) is smeared out over an
interval $\Delta T$ according to \cite{rischke}
\be
s=\frac{4\pi^2}{90}T^3 \frac{d_Q-d_H}{2}
\left[\frac{d_Q+d_H}{d_Q-d_H}+\tanh\left(\frac{T-T_M}{\Delta T}\right)\right]
\Punkt \label{secndeos}
\ee
In this scheme, the time evolution of the temperature has to be
calculated numerically using Eq.~(\ref{bjent}). The result is a smooth
curve, which for $\Delta T\to 0$ approaches that of the first order
transition, as can be seen from the dashed ($\Delta T=10~\mbox{MeV}$) and
dot--dashed ($\Delta T=1~\mbox{MeV}$) curves in Fig.~\ref{temper}.
However, this scenario displays the qualitative difference to the
first order scenario, that it does not develop a mixed phase. Note that for
$T_0-T_M\gg\Delta T$, the phase transition is reached at time
\be
\tau_1=\tau_0 \left(\frac{T_0}{T_M}\right)^3 \frac{2d_Q}{d_Q+d_H}
\Punkt
\ee
For the parameter set of Fig.~\ref{temper}, this yields
$\tau_1=3.35~\mbox{fm}/c$. This is very small compared to the
end time of the mixed phase in the first order scenario.

In contrast to Eq.~(\ref{firstspect}), the photon spectrum is now
given by
\be
\frac{dN_{\gamma\gamma}}{dMd\eta} = \pi R_A^2
\int_{\tau_0}^{\tau_1} d\tau \, \tau \frac{dN_{\gamma\gamma}}{d^4xdM}
\Punkt
\ee
As Eq.~(\ref{firstspect}), the integration is cut off at the phase
transition and thus contains only contributions of a pure quark phase.

\subsection{Photon Spectra}
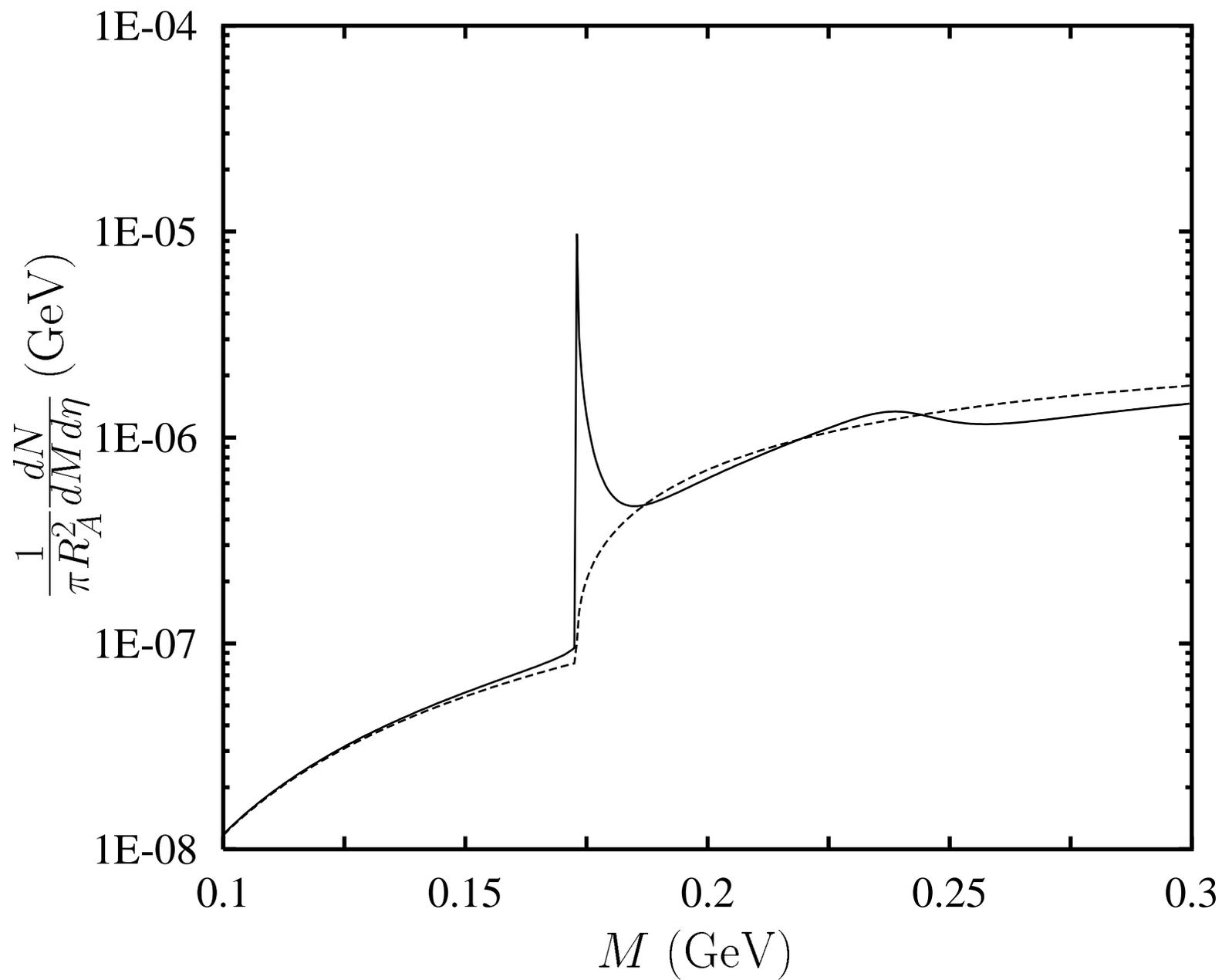
\begin{figure}
\caption[]{Comparison of photon spectra containing resonance contributions
           (solid line) and photon spectra without resonance
           contributions (dashed line) in the case of a first order
           transition at central rapidity as a function of the invariant
           pair mass $M$.}
\label{result1}
\end{figure}

The physical photon spectra at central rapidity for the case of a first
order phase transition are shown in Fig.~\ref{result1} as a function of
the invariant pair mass. The dashed line gives the photon spectra
without resonance contributions. Since the system stays at temperature
$T=T_M$ for a finite time, the spectra develop a cusp at
$M=2m_q(T_M)$.  The rise in the photon yield at larger pair masses
reflects the contribution of the mixed phase.  The photon spectra
containing resonance contributions is given by the solid line in
Fig.~\ref{result1}. Here the influence of the mixed phase is more
dramatic: Due to the second term of Eq.~(\ref{firstspect}), the
threshold singularity in Eq.~(\ref{nkri}) immediately carries over to
the physical spectra and one thus observes a great enhancement of
photon pairs at the invariant mass $M=2m_q(T_M)=m_\pi(T_M)$. For our
parameter set, this occurs at $M=173$~MeV. Note that this peak is
in principle separated from the hadronic contributions due to $\pi^0$
decay ($M=135$~MeV) and due to $\pi^+\pi^-$ annihilation
($M \ge 279$~MeV).  However, we suspect that this medium effect might
hardly be visible in experiment since it is not clear if the
enhancement of the pion mass is sufficient to separate the critical
scattering peak from the huge peak in the photon spectra stemming from
the decay of ``cold pions'' originating after hadronization. A
quantitative estimate of the number of photon pairs originating from
$\pi^0$ decay can be made from the relation \cite{kohl}
\be
\frac{1}{\pi R_A^2} \frac{dN_{\pi^0}}{d\eta}
= 1.5\, \tau_0 T_0^3 = 3.1\, \mbox{fm}^{-2} \Punkt \label{pinum}
\ee
The total number of photon pairs contained in the solid curve of
Fig.~\ref{result1} is
\be
\frac{1}{\pi R_A^2} \frac{dN_{\gamma\gamma}}{d\eta}
= 4.0 \times 10^{-5}\, \mbox{fm}^{-2} \Komma
\ee
which is of the order of a factor $\alpha_{\rm em}^2$ less than the value
given in Eq.~(\ref{pinum}). The number of photon pairs contained in the
Mott peak is
\be
\frac{1}{\pi R_A^2} \frac{dN_{\gamma\gamma}}{d\eta}
= 3.8 \times 10^{-7}\, \mbox{fm}^{-2} \Punkt
\ee
The Mott peak thus contains roughly 1\% of the total number of
photons stemming from $q\bar q$ annihilation.

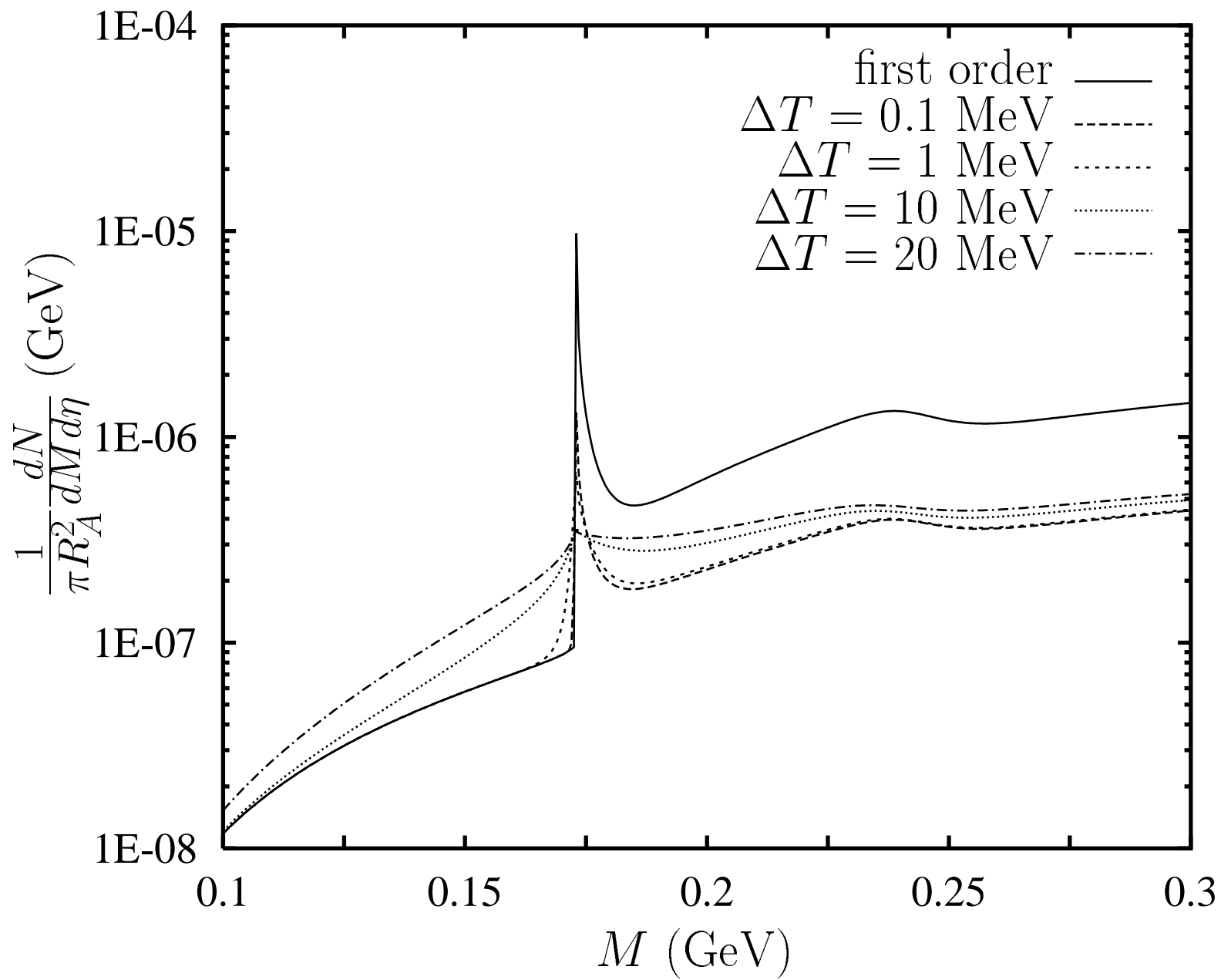
\begin{figure}
\caption[]{Comparison of photon spectra containing resonance contributions
           for a first order phase transition (solid line) and for a
           second order phase transition with various values of $\Delta
           T$ at central rapidity as a function of the invariant pair
           mass $M$.}
\label{result2}
\end{figure}

In the case of a second order phase transition, the system does not
stay at a fixed temperature for a finite time. Thus the Mott peak
visible in Fig.~\ref{result1} should disappear if the order of the
phase transition is weakened. Figure~\ref{result2} shows the effect
of the parameter $\Delta T$ on the photon spectra. Here the case of
the first order phase transition is shown as the solid line. The
other lines show the spectra in the case of a second order phase
transition for various values of $\Delta T$. One recognizes that
one still obtains a strongly enhanced photon yield at $M=m_\pi(T_M)$
for very small values of $\Delta T$. For larger values of
$\Delta T$, however, this peak gets flattened. At $\Delta T=20$~MeV,
only a weak cusp is visible in the two-photon spectra from a quark-meson
plasma.

\section{Summary and Conclusions}
\label{end}
We have investigated the influence of resonance contributions on the
process $q\bar q\to\gamma\gamma$ in a quark meson plasma at
temperatures larger than the Mott temperature.  We have employed a
chiral quark model as a microscopic approach to a consistent
description of the chiral phase transition and the Mott effect of the
pion as a $q\bar q$ bound state. We have found that at the pion Mott
temperature $T_M$, which corresponds to the chiral transition
temperature $T_\chi$ present for $m_{0q}=0$, there occurs a narrow peak
in the two photon spectrum at the invariant mass $M=m_\pi(T_M)\approx
173$~MeV. Note that lattice results give $T_\chi=T_{\rm deconfinement}$
within the present accuracy \cite{laer}.  It seems thus natural to
identify the Mott point with the deconfinement temperature.

In order to estimate the role of this critical scattering effect on
the photon spectra emitted from an expanding quark plasma, as produced
e.\,g. in heavy ion collisions, we have used the hydrodynamical Bjorken
scenario. The expansion dynamics fixes the temperature near $T=T_M$ for
a finite time interval, depending on the parameter $\Delta T$ of
Eq.~(\ref{secndeos}), which is a measure for the smoothness of the
transition. In the case of a first order transition this leads to a
divergence in the photon spectra which is, however, only of the order
$(M-m_\pi(T_M))^{-1/2}$, so that the total number of photons emitted is
finite. In the case of a second order phase transition, the divergence
disappears. However, since the temperature stays {\em near} $T_M$ for
finite time, the critical scattering peak remains visible.  The
threshold singularity of the cross section, which is responsible for
the critical scattering effect, may be weakened due to higher order
corrections in $1/N_c$. We nevertheless suspect that the cross section
stays large at threshold.

In the present work we have limited ourselves to the investigation of
the photon pair production in the quark phase and did not take into
account the contribution of the later expansion stage. This calculation
is considered as a first step towards a more complete analysis of in
medium effects on photon production in heavy ion collisions within
chiral quark models. The hadronic tail of the evolution of a hot quark
plasma phase also gives important contributions. A study of the
processes in a pion gas, as $\pi^0\to \gamma \gamma$ and
$\pi^+\pi^-\to  \gamma \gamma$, is in progress \cite{volku}. The
applicability of quark models without confinement in the hadronic phase
is limited to those kinematical situations, where unphysical quark
production thresholds do not play a role. A first step towards the
development of chiral confining quark models at finite temperature has
been undertaken in Ref.~\cite{confin}.  Also in these more realistic
models the Mott effect occurs and we expect its manifestation in the
two photon spectra.

\section*{Acknowledgments}
We wish to thank J.~H\"ufner, S.\,P.~Klevansky and M.\,K.~Volkov for
illuminating discussions and a careful reading of the manuscript. This
work has been supported in part by the Deutsche Forschungsgemeinschaft
under contracts no. 436 RUS 17/168/95 and Hu 233/4-4, and the Federal
Ministry for Education and Research under contract no. 06 HD 742.

\begin{appendix}
\section{Meson to Photon Decay Vertices} \label{appvertex}
\subsection{$\sigma\to\gamma\gamma$}
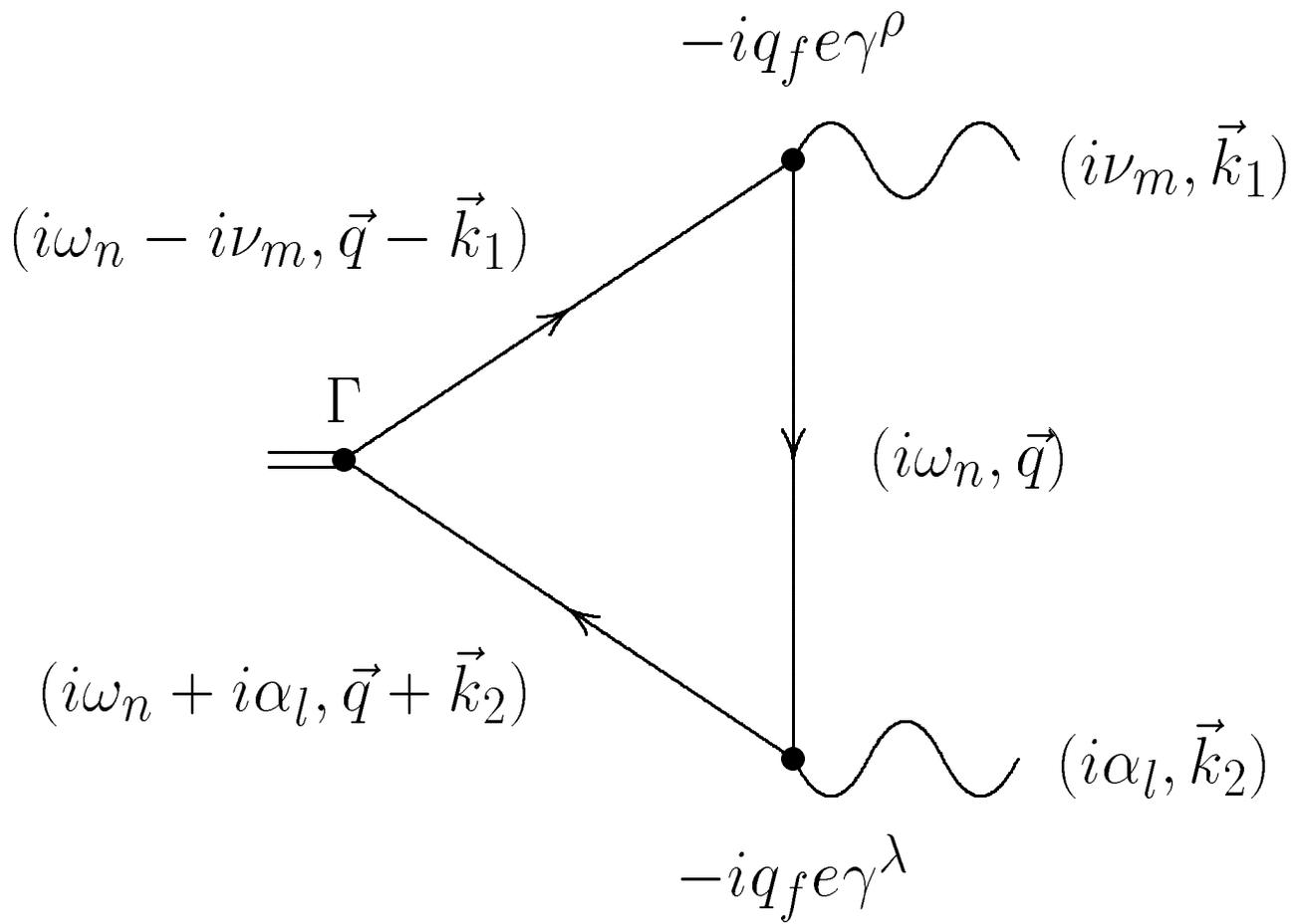
\begin{figure}
\caption[]{Feynman diagram for the meson to photon decay vertex. Here
           one has to set $\Gamma=1$ for $\sigma\to\gamma\gamma$ and
           $\Gamma=i\gamma_5\tau^3$ for $\pi^0\to\gamma\gamma$. A
           second diagram arises by the substitution
           $k_1\leftrightarrow k_2$ and $\rho \leftrightarrow \lambda$.}
\label{decgraph}
\end{figure}

The generic Feynman graph for the vertex $\sigma\to\gamma\gamma$ is
given by Fig.~\ref{decgraph} with $\Gamma=1$. The contribution due to
this diagram is
\be
-iA^{\rho\lambda}_{\sigma\to\gamma\gamma} =
- \frac{i}{\beta}\sum_n \int \frac{d^3q}{(2\pi)^3}
\mbox{Tr}\left[iS_f(q) (-i\gamma^\rho q_f e)
               iS_f(q - k_1) iS_f(q + k_2)
               (-i\gamma^\lambda q_f e) \right] \label{eqa1}
\ee
where we have introduced the symbolic four vector notation
\be
q = (i\omega_n,\vec q) \qquad k_1 = (i\nu_m, \vec k_1)
\qquad k_2=(i\alpha_l, \vec k_2) \Punkt
\ee
In Eq.~(\ref{eqa1}), $\omega_n$ is a fermionic Matsubara frequency,
whereas $\nu_m$
and $\alpha_l$ are bosonic Matsubara frequencies belonging to the
outgoing photons. It is understood that $i\nu_m$ and $i\alpha_l$
have to be analytically continued to real values at the end of
the calculation.  The symbol Tr denotes the traces over spin, color and
flavor degrees of freedom. Taking the trace over color and flavor
yields a factor
\be
N_c \left(q_u^2+q_d^2\right) = \frac{5}{9} N_c \Komma
\ee
so that we have
\be
-iA^{\rho\lambda}_{\sigma\to\gamma\gamma} =
\frac{5}{9} \frac{N_c e^2}{\beta}\sum_n \int \frac{d^3q}{(2\pi)^3}
\frac{\mbox{tr}\left[
(q \Slash + m_q)\gamma^\rho(q \Slash - k_1 \SLAsh + m_q)
(q \Slash + k_2 \SLAsh + m_q) \gamma^\lambda \right]}
{[q^2-m_q^2][(q-k_1)^2 - m_q^2][(q+k_2)^2-m_q^2]} \Komma
\ee
where now the symbol tr denotes the spinor trace only. After taking 
the trace and noting that terms proportional to $k_1^\rho$ and
$k_2^\lambda$ do not contribute to the final result due to the
transversality of the photon, one has
\be
-iA^{\rho\lambda}_{\sigma\to\gamma\gamma} =
\frac{20}{9} \frac{N_c m_q e^2}{\beta}\sum_n \int \frac{d^3q}{(2\pi)^3}
\frac{(k_1^\lambda k_2^\rho - g^{\rho\lambda} k_1k_2) - g^{\rho\lambda}
(q^2-m_q^2) + 4 q^\rho q^\lambda}
{[q^2-m_q^2][(q-k_1)^2 - m_q^2][(q+k_2)^2-m_q^2]} \Punkt
\ee
Decomposing this integral and performing the analytical continuation
leads to
\be
-iA^{\rho\lambda}_{\sigma\to\gamma\gamma} =
\frac{5m_qN_c\alpha_{\rm em}}{9\pi}
(k_1^\lambda k_2^\rho - g^{\rho\lambda} k_1k_2)
\left(1-\frac{2m_q^2}{q_1q_2}\right) C_0(k_1, -k_2) \label{fidi}
\ee
where $\alpha_{\rm em}=1/137$ is the electromagnetic fine structure
constant and
\be
C_0(k_1, -k_2) = \frac{16\pi^2}{\beta} \sum_n
\int \frac{d^3q}{(2\pi)^3}
\frac{1}{[q^2-m_q^2][(q-k_1)^2 - m_q^2][(q+k_2)^2-m_q^2]} \Punkt
\ee
In the kinematical situation required throughout this work, $\vec k_1 =
- \vec k_2$, $k_1^0=k_2^0=|\vec k_1| = |\vec k_2| = \sqrt{s}/2$, this
integral can be evaluated to be \cite{loopies}
\bea
C_0(k_1, -k_2) &=& -2 {\cal P} \! \! \! \int_0^\Lambda dp
\frac{p\tanh(\beta E/2)}{E^2(s-4E^2)} \log\left|\frac{E-p}{E+p}\right|
\\ \nonumber &+&
i\frac{\pi}{s} \tanh(\beta \sqrt{s}/4)
\log\left|\frac{\sqrt{s} - \sqrt{s-4m_q^2}} {\sqrt{s} + \sqrt{s-4m_q^2}}
\right|
\eea
and is thus a function of $s$ only.
Beside (\ref{fidi}), there is a second
contribution to the vertex from the crossed diagram of
Fig.~\ref{decgraph}, which arises by exchanging $k_1\leftrightarrow
k_2$ and $\rho \leftrightarrow \lambda$. The contribution of this
diagram is found to be equal to that of the first diagram.  The
function $a(s)$ in Eq.~(\ref{s1chan}) may be thus written as
\be
-ia(s) = \frac{10m_qN_c\alpha_{\rm em}}{9\pi}
\left(1-\frac{4m_q^2}{s}\right) C_0(k_1, -k_2) \Punkt
\ee
This result is in agreement with the result of Ref.~\cite{blihi}.

\subsection{$\pi^0\to\gamma\gamma$}
The vertex $\pi^0\to\gamma\gamma$ can again be computed from
Fig.~\ref{decgraph}, but now one has to take $\Gamma=i\gamma_5\tau^3$.
This leads to
\bea
-iA^{\rho\lambda}_{\pi^0\to\gamma\gamma} &=&
- \frac{i}{\beta}\sum_n \int \frac{d^3q}{(2\pi)^3}
\\ \nonumber &\times&
\mbox{Tr}\left[iS_f(q) (-i\gamma^\rho q_f e)
               iS_f(q - k_1) (i\gamma_5\tau^3) iS_f(q + k_2)
               (-i\gamma^\lambda q_f e) \right]
\eea
in the same notation as in Eq.~(\ref{eqa1}). The trace over color and
flavor degrees of freedom yields now
\be
N_c\left(q_u^2-q_d^2\right) = \frac{1}{3} N_c
\ee
and one has
\be
-iA^{\rho\lambda}_{\pi^0\to\gamma\gamma} =
\frac{i}{3} \frac{N_c e^2}{\beta}\sum_n \int \frac{d^3q}{(2\pi)^3}
\frac{\mbox{tr}\left[
(q \Slash + m_q)\gamma^\rho(q \Slash - k_1 \SLAsh + m_q) \gamma_5
(q \Slash + k_2 \SLAsh + m_q) \gamma^\lambda \right]}
{[q^2-m_q^2][(q-k_1)^2 - m_q^2][(q+k_2)^2-m_q^2]} \Punkt
\ee
The only terms of the spinor trace which give a contribution are those,
which involve $\gamma_5$ and four other gamma matrices. One obtains
\be
-iA^{\rho\lambda}_{\pi^0\to\gamma\gamma} =
\frac{\alpha_{\rm em}m_qN_c}{3\pi}
\varepsilon^{\rho\lambda}_{\phantom{\rho\lambda}\gamma\delta}
k_1^\gamma k_2^\delta C_0(k_1, -k_2)
\ee
and the same contribution from the crossed graph. In total,
one has for $b(s)$ required in Eq.~(\ref{s2chan})
\be
-ib(s) = \frac{2\alpha_{\rm em}m_qN_c}{3\pi} C_0(k_1, -k_2) \Punkt
\ee
This agrees with the result of Ref.~\cite{sapho}.

\section{Squared Transition Amplitudes} \label{appmatel}
To compute the integrated cross section, one need the square of the
transition amplitude, averaged over incoming and summed over
outgoing states,
\be
\frac{1}{4N_c^2}\sum_{s,c}\left| {\cal M}_t + {\cal M}_u +
{\cal M}_{s,\sigma} + {\cal M}_{s,\pi}\right|^2 \Punkt \label{eqb1}
\ee
We will quote our results for the single terms in Eq.~(\ref{eqb1}),
beginning with the $t$- and $u$-channel diagrams. For the squares of
these diagrams, one has
\bea
\frac{1}{4N_c^2}\sum_{s,c}
\left|{\cal M}_t\right|^2 &=& \frac{32 \pi^2 q_f^4 \alpha_{\rm em}^2}
{N_c(t-m_q^2)^2} \left[ (m_q^2-t)(s-2m_q^2) - (t+m_q^2)^2\right] \\
\frac{1}{4N_c^2}\sum_{s,c}
\left|{\cal M}_u\right|^2 &=& \frac{32 \pi^2 q_f^4 \alpha_{\rm em}^2}
{N_c(u-m_q^2)^2} \left[ (m_q^2-u)(s-2m_q^2) - (u+m_q^2)^2\right] \Komma
\eea
whereas the interference term yields
\be
\frac{1}{4N_c^2}\sum_{s,c}
{\cal M}_t{\cal M}_u^* = \frac{32 \pi^2 q_f^4 \alpha_{\rm em}^2}
{N_c(t-m_q^2)(u-m_q^2)} m_q^2(s-4m_q^2) \Punkt
\ee
Performing the angular integration, one obtains
\bea
\int_{t_{\rm min}}^{t_{\rm max}} dt \frac{1}{4N_c}^2 \left|
{\cal M}_t + {\cal M}_u \right|^2 &=&
\frac{32 \pi^2 q_f^4 \alpha_{\rm em}^2}{N_c s}
\\ \nonumber &\times&
\Bigg[(s^2 + 4m_q^2 s - 8 m_q^4)
\log\frac{\sqrt{s}+\sqrt{s-4m_q^2}}{\sqrt{s}-\sqrt{s-4m_q^2}}
\\ \nonumber & &
-(s+4m_q^4) \sqrt{s(s-4m_q^2)} \Bigg]
\eea
and from this the result for the integrated cross section {\em without}
resonances \cite{lanlif}.

For the squares of the $s$--channel diagrams, one has
\be
\frac{1}{4N_c^2}\sum_{s,c} \left|{\cal M}_{s,\sigma}\right|^2 =
\frac{1}{4N_c} s^2(s-4m_q^2) \left|a(s){\cal D}_\sigma(\sqrt{s},\vec 0)
\right|^2
\ee and \be
\frac{1}{4N_c^2}\sum_{s,c} \left|{\cal M}_{s,\pi}\right|^2 =
\frac{1}{4N_c} s^3 \left|b(s){\cal D}_\pi(\sqrt{s},\vec 0) \right|^2
\Punkt
\ee
The interference term of the scalar and pseudoscalar $s$--channels
vanishes. For the interference terms of the $s$--channel amplitudes
with the direct amplitudes, one obtains
\bea
\frac{1}{4N_c^2}\sum_{s,c} {\cal M}_{s,\sigma}{\cal M}_t^* &=&
\frac{2\pi q_f^2\alpha_{\rm em}m_q}{N_c}
\frac{\left[-ia(s){\cal D}_\sigma(\sqrt{s},\vec 0)\right]} {t-m_q^2}
\\ \nonumber & & \hspace{1cm} \times
\left[s(s-4m_q^2)-(t-u)(t-m_q^2)\right] \\
\frac{1}{4N_c^2}\sum_{s,c} {\cal M}_{s,\sigma}{\cal M}_u^* &=&
\frac{2\pi q_f^2\alpha_{\rm em}m_q}{N_c}
\frac{\left[-ia(s){\cal D}_\sigma(\sqrt{s},\vec 0)\right]} {u-m_q^2}
\\ \nonumber & & \hspace{1cm} \times
\left[s(s-4m_q^2)-(u-t)(u-m_q^2)\right]
\eea
for the sigma exchange and
\bea
\frac{1}{4N_c^2}\sum_{s,c} {\cal M}_{s,\pi}{\cal M}_t^* &=&
\frac{2\pi \tau^3_{ff} q_f^2\alpha_{\rm em}m_q}{N_c}
\frac{\left[-ib(s){\cal D}_\pi(\sqrt{s},\vec 0)\right]} {t-m_q^2} s^2 \\
\frac{1}{4N_c^2}\sum_{s,c} {\cal M}_{s,\pi}{\cal M}_u^* &=&
\frac{2\pi \tau^3_{ff} q_f^2\alpha_{\rm em}m_q}{N_c}
\frac{\left[-ib(s){\cal D}_\pi(\sqrt{s},\vec 0)\right]} {u-m_q^2} s^2
\eea
for the pion exchange. Integrating over $t$ leads to the results
\bea
\int_{t_{\rm min}}^{t_{\rm max}} dt \frac{1}{4N_c^2} \left|
{\cal M}_{s,\sigma} + {\cal M}_{s,\pi} \right|^2 &=&
\frac{s^2\sqrt{s(s-4m_q^2)}}{8N_c}\left[(s-4m_q^2)
\left|a(s){\cal D}_\sigma(\sqrt{s},\vec  0) \right|^2 \right.
\nonumber \\ & & \qquad + \left.
s \left|b(s){\cal D}_\pi(\sqrt{s},\vec  0) \right|^2 \right]
\label{litschi}
\eea
and
\bea
& &
\int_{t_{\rm min}}^{t_{\rm max}} dt \frac{1}{4N_c^2} \left(
{\cal M}_{s,\sigma} + {\cal M}_{s,\pi} \right) \left(
{\cal M}_t + {\cal M}_u \right)^* =
\\ \nonumber & & \qquad
-\frac{2\pi m_q q_f^2 \alpha_{\rm em} s}{N_c}
\log\frac{\sqrt{s}+\sqrt{s-4m_q^2}}{\sqrt{s}-\sqrt{s-4m_q^2}}
\left[(s-4m_q^2) \left(-ia(s) {\cal D}_\sigma(\sqrt{s}, \vec 0) \right) \right.
\\ \nonumber & & \qquad \left.
+ s \tau^3_{ff} \left(-ib(s) {\cal D}_\pi(\sqrt{s}, \vec 0)\right) \right]
\Punkt
\eea
From this, one immediately obtains the integrated cross section via
Eq.~(\ref{sigtot}).

\end{appendix}

\end{document}